\documentclass[11pt,a4paper]{article}
\pdfoutput=1
\usepackage{jheppub}
\usepackage{bbm}
\usepackage{bm}
\usepackage{footmisc}
\usepackage{bbold}
\usepackage{graphicx}
\usepackage[T1]{fontenc}
\usepackage[utf8]{inputenc}
\usepackage[english]{babel}
\usepackage[usenames,dvipsnames]{xcolor}
\usepackage[normalem]{ulem}

\usepackage{epsfig}
\usepackage{amsmath,amssymb,amsfonts}
\usepackage{mathrsfs}
\usepackage{breqn}
\usepackage[bbgreekl]{mathbbol}
\usepackage{slashed}
\usepackage{verbatim}

\newcommand{\UV}{{\small UV}}

\newcommand{\FRG}{{\small FRG}}
\newcommand{\RG}{{\small RG}}
\newcommand{\CDT}{{\small CDT}}
\newcommand{\EDT}{{\small EDT}}
\newcommand{\ADM}{{\small ADM}}
\newcommand{\eg}{{\textit{e.g.}}}
\newcommand{\ie}{{\textit{i.e.}}}
\newcommand{\EFT}{{\small EFT}}

\newcommand{\hmat}{\ensuremath{\mathbbm h}}

\begin{document}

\title{Configuration space for quantum gravity in a locally regularized path integral}

\author[a]{\href{https://orcid.org/0000-0001-6700-6501}{\protect \includegraphics[scale=.07]{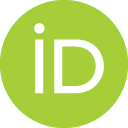}} Benjamin Knorr,} 
\author[a]{\href{https://orcid.org/0000-0001-7789-344X}{\protect \includegraphics[scale=.07]{ORCIDiD_icon128x128.png}} Alessia Platania}
\author[a]{and \href{https://orcid.org/0000-0002-0778-4800}{\protect \includegraphics[scale=.07]{ORCIDiD_icon128x128.png}} Marc Schiffer} 

\affiliation[a]{Perimeter Institute for Theoretical Physics, 31 Caroline Street North, Waterloo, ON N2L 2Y5, Canada}

\emailAdd{bknorr@perimeterinstitute.ca}
\emailAdd{aplatania@perimeterinstitute.ca}
\emailAdd{mschiffer@perimeterinstitute.ca}

\abstract{We discuss some aspects of the metric configuration space in quantum gravity in the background field formalism. We give a necessary and sufficient condition for the parameterization of Euclidean metric fluctuations such that i) the signature of the metric is preserved in all configurations that enter the gravitational path integral, and ii) the parameterization provides a bijective map between full Euclidean metrics and metric fluctuations about a fixed background. 
For the case of foliatable manifolds, we show how to parameterize fluctuations in order to preserve foliatability of all configurations. 
Moreover, we show explicitly that preserving the signature on the configuration space for the Lorentzian quantum gravitational path integral is most conveniently achieved by inequality constraints. We discuss the implementation of these inequality constraints in a non-perturbative renormalization group setup.}


\maketitle

\flushbottom

\section{Introduction}

The arguably most conservative framework to develop a quantum theory of gravity is to quantize the metric within a path integral formalism \cite{tHooft:1974toh, Gibbons:1976ue, DeWitt:1967ub, Mottola:1995sj}.
In order to define the quantum gravitational path integral, a configuration space must be chosen. For instance, one could restrict the spacetime dimensionality and topology as well as the signature of the metric, \eg{}, summing over Lorentzian spacetimes with trivial spacetime topology only. One might 
further restrict the gravitational path integral, for example, by summing over  globally hyperbolic spacetimes only.

These conditions select a subset of the space of all metrics, constraining the integration  domain in the path integral. 
Imposing such constraints is complicated by the fact that the path integral also requires a regularization. The unregularized path integral is ill-defined, and therefore it is crucial to understand how constraints and regularization can both be implemented in the path integral at the same time.

Defining a regularization requires to thin out the configuration space appropriately, such that in the limit of vanishing thinning parameter, the full configuration space is recovered. The thinning should be implemented in such a way that at finite thinning parameter, the resulting path integral is finite. Further, it is desirable that the thinning is physically meaningful, in that the selected configurations should be physically relevant. The latter property is expected to be key to obtain robust approximations.

In the context of a quantum field theory on a flat background the regularization can be implemented, after Wick-rotating to Euclidean signature, via an ultraviolet cutoff in momentum space.\footnote{A cutoff on the eigenmodes of the covariant Laplacian with respect to the background metric is a natural generalization to the case of a non-trivial background metric which admits a Wick rotation, see, \eg{}, \cite{Baldazzi:2018mtl} for a discussion of the latter requirement.} This is tied to a local notion of coarse-graining, where locality refers to a background metric.

Regularizing the quantum gravitational path integral is a more subtle issue than it is in ``standard'' quantum field theories. In continuum, metric\footnote{Similar arguments may hold in cases where the fundamental degrees of freedom are not described by the metric, but by other fields, such as the vielbein, torsion, holonomies, or the affine connection.} approaches to quantum gravity, the thinning procedure comes at considerable extra cost, as it typically requires the introduction of a \emph{background metric}, for potential alternatives see \cite{Morris:2016nda, Falls:2020tmj}. A background metric is not a natural notion in quantum gravity, unless one explores only the strictly perturbative regime. Otherwise, all configurations of spacetime should be treated on the same footing, thus preserving background independence.
In a manifestly background-independent path integral over metrics there is no notion of locality, and thus local coarse-graining requires the introduction of a distinguished metric using background field techniques \cite{Abbott:1980hw, Reuter:1996cp}.\footnote{Avoiding the introduction of a background metric leads to pregeometric settings~\cite{Oriti:2018dsg, Wetterich:2021ywr, Wetterich:2021hru, Pedraza:2021fgp, Oriti:2021oux}, where the use of non-metric degrees of freedom allows to coarse-grain in a background independent way, see, \eg{}, \cite{Rivasseau:2011hm, Eichhorn:2018phj, Steinhaus:2020lgb}. As a drawback, the connection to continuous geometries becomes a highly non-trivial one. Thus, restrictions on the configuration space of geometries can become highly challenging to interpret in this pregeometric context.}

Therefore we are led to the challenge of restricting the configuration space of metric fluctuations, within the background field formalism in such a way that the regularization is consistent with imposing the constraints, \eg{}, on spacetime signature and topology.

For our discussion it is useful to distinguish two different types of constraints. The first type are ``equality constraints'' where allowed configurations have to satisfy a strict equality.  These are the standard constraints encountered in classical mechanics (\eg{}, the mass of a pendulum is constrained to be on a circle of a given radius, the string length). In a quantum field theory setup, such constraints can be implemented using Dirac delta distributions. To this end, one can perform a constraint analysis \`a la Dirac \cite{Dirac:1950pj, Dirac:1958jc}, determine the full set of second class constraints, and implement them at the level of the path integral using the Senjanovic method  \cite{Senjanovic:1976br}. The second type of constraints, and the most relevant for the present discussion, are ``inequality constraints'' where allowed configurations have to satisfy an inequality. The equivalent in classical mechanics would be, \eg{}, a ball bouncing off a hard surface (see, \eg, \cite{Kleinert} for an implementation in the quantum-mechanical path integral). A prominent example for an inequality constraint in gravity is the restriction of metric fluctuations such that the signature of the full metric stays fixed. It is  a main  goal of this article to understand how this type of constraints can be implemented in the gravitational path integral, see also~\cite{Dittrich:2004bn, Bahr:2009ku, Dittrich:2013jaa, Asante:2020qpa, Asante:2021zzh}.

This paper is organized as follows:
In section \ref{sec:pathintegral} we discuss some preliminaries on the path integral, illustrate the problem and introduce our notation. In section \ref{sec:Eucl-param}, we first highlight our strategy to best implement inequality constraints on the space of field configurations: finding a suitable parameterization of the fluctuations such that the path integral can be taken over the unconstrained fields. We then discuss the construction of signature-preserving metric parameterizations in Euclidean quantum gravity. We show that there are infinitely many possible parameterizations which define bijective maps from the space of real symmetric fluctuation fields to Euclidean metrics of a given topology, given a background metric of the same topology. As has been emphasized before \cite{Nink:2014yya, Percacci:2015wwa}, the linear parameterization is not one of them, while the exponential one \cite{Kawai:1993fq, Eichhorn:2013xr, Nink:2014yya, Percacci:2015wwa} is. In section \ref{sec:foliation}, we then explore how to impose the constraint that a foliation exists in a consistent and covariant way, extending previous work in~\cite{Knorr:2018fdu}.\footnote{At the level of a classical Lagrangian or Hamiltonian, the correct implementation of a constraint such as a foliation structure requires to first identify all constraints via the Dirac analysis. At the level of an off-shell path integral however, it is not clear that the Dirac analysis is necessary. In particular, one could first aim at resumming all quantum fluctuations, and determine the full set of constraints at the level of the effective Lagrangian only.} In particular, we show that a foliation structure, where all fluctuations preserve the signature of the full metric, can be implemented by a suitable choice of parameterization, without adding additional constraints to the path integral. 
We turn our attention to Lorentzian signature in section~\ref{sec:Lorentzian} and discover that beyond the perturbative setting there are severe obstructions to the construction of a bijective map between the space of real symmetric fluctuation fields and the space of Lorentzian metrics, generalizing the discussion of the exponential parameterization in \cite{Demmel:2015zfa}. In particular, we present an explicit example showing why the Euclidean strategy is doomed to fail in the Lorentzian case. This suggests that using the background field method in the Lorentzian setup requires a direct implementation of an inequality constraint. We illustrate how this can be done within the Functional Renormalization Group (\FRG{}) framework in section \ref{sec:ineq-constraints} in the case of a scalar field theory subject to an inequality constraint, and we derive a modified version of the \FRG{} equation \cite{Wetterich:1992yh, Morris:1993qb, Ellwanger:1993mw, Reuter:1996cp} which takes the inequality constraint into account.  Finally, we discuss and summarize our findings in section \ref{sec:conclusions}.

\section{The path integral over metrics}\label{sec:pathintegral}

\begin{figure}
	\centering
	\includegraphics[width=\linewidth, clip=true,trim=0cm 3.5cm 0cm 2.5cm]{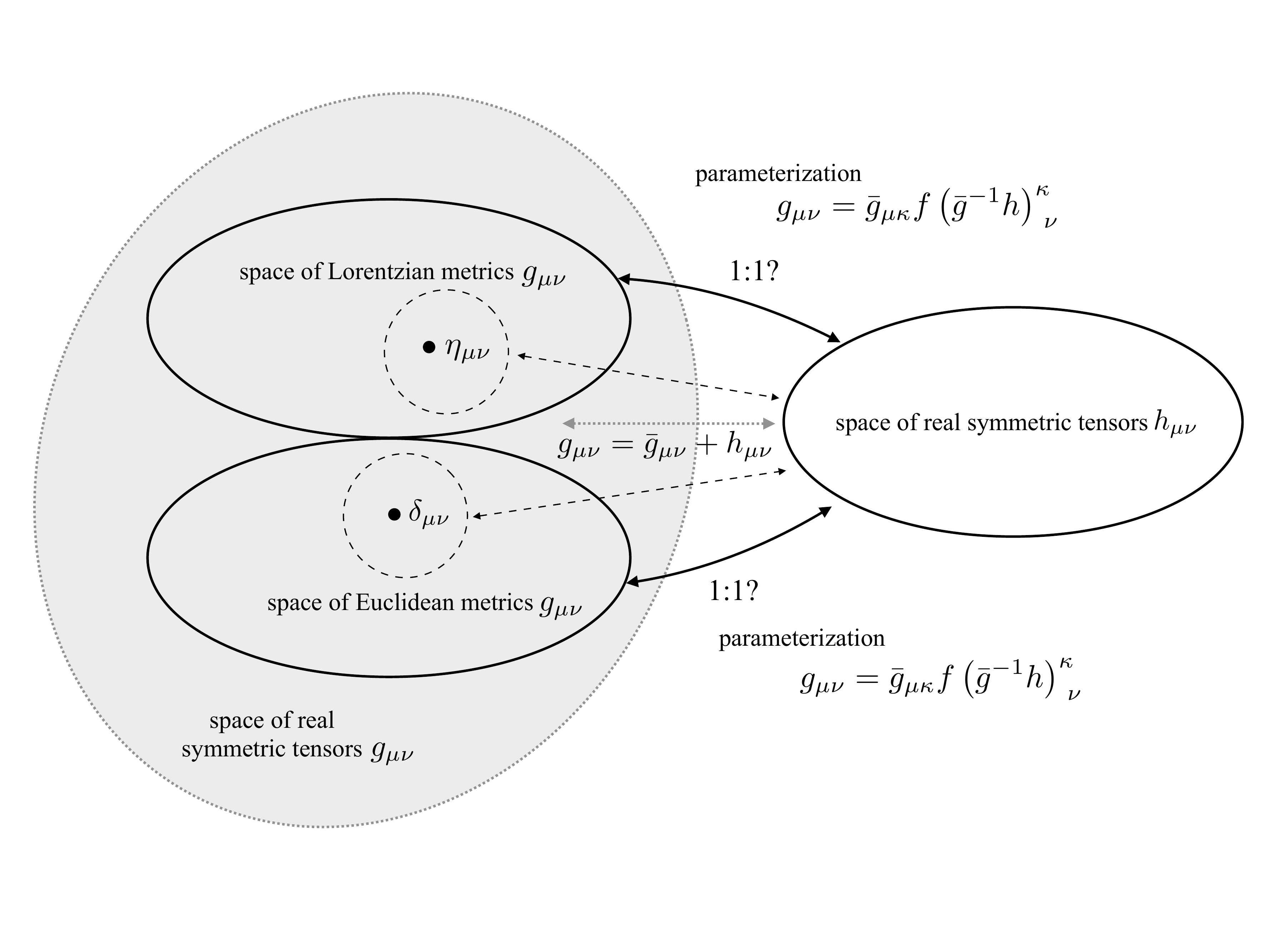}
	\caption{\label{fig:illustration1}We illustrate the space of Lorentzian metrics and the space of Euclidean metrics. It is the key question of this work whether a one-to-one map exists from the space of real symmetric tensors $h_{\mu\nu}$ to the spaces of metrics of fixed signature. It is known that the linear parameterization maps to a set that contains both Lorentzian and Euclidean metrics (light grey region), unless one works in perturbation theory (indicated by dashed lines).}
\end{figure}

In this section we motivate the necessity of  finding a bijective map between a background metric and the full metric on the level of the gravitational path integral.

Our discussion focuses on approaches to quantum gravity that are based on a path integral over spacetime metric configurations. One difficulty in this approach is that the gravitational path integral
	\begin{equation}
	\mathcal{Z} = \int \mathcal{D}g_{\mu\nu} \, e^{i S[g_{\mu\nu}]} \, ,
	\end{equation} 
where $S$ is the gravitational action, is ill-defined, as it misses a regularization.
Lattice methods to evaluate the path integral of quantum gravity, such as Euclidean or Causal dynamical triangulations (\EDT{}/\CDT{}) \cite{Laiho:2017htj, Loll:2019rdj}, implement this regularization via a discretization of spacetime. Physical quantities are then extracted in the continuum limit, if it exists.
In continuum metric formulations, the ill-defined path integral over the metric $g_{\mu\nu}$ is replaced by one over metric fluctuations $h_{\mu\nu}$ around a background metric $\bar{g}_{\mu\nu}$, where a suitable regularization is implemented, namely:
\begin{equation}
\label{eq:PIs}
\mathcal{Z}_R = \int\mathcal{D}_R h_{\mu\nu}\,e^{i S[\bar{g}_{\mu\nu}; h_{\mu\nu}]}\,,
\end{equation}
where we indicated that the measure also includes a regularization. A generic difficulty in this formulation of quantum gravity is how to make sure that quantum fluctuations $h_{\mu\nu}$ of the metric $g_{\mu\nu}$ about a background metric $\bar{g}_{\mu\nu}$ extend only over those configurations that one wants to integrate over. Here, we work under the assumption that only metrics of a fixed signature (either Euclidean or Lorentzian) should contribute to the gravitational path integral.\footnote{Other possibilities can be considered, see, \eg{}, \cite{Ellis:1991st, Dray:1996dc, Bojowald:2015gra}.} Our work aims at finding conditions to construct parameterizations of the metric fluctuations which realize a one-to-one map between the space of real symmetric tensors $h_{\mu\nu}$ and the spaces of metrics of fixed (Euclidean or Lorentzian) signature. Our idea is summarized in figure \ref{fig:illustration1}.

A simpler version of the full path integral \eqref{eq:PIs} is typically considered in  perturbative setups, where the metric fluctuations $h_{\mu\nu}$ are assumed to be small perturbations around a flat background, \ie{}, $g_{\mu\nu}=\eta_{\mu\nu}+h_{\mu\nu}$. In this effective field theoretical (\EFT{}) approach to quantum gravity, see, \eg{} \cite{Donoghue:1993eb, Donoghue:1994dn}, the regularization of the path integral is implemented via an explicit \UV{} cutoff, given by the Planck scale. Below the cutoff, a perturbative treatment of quantum gravity is viable. The \EFT{} only breaks down above the cutoff scale where the theory becomes strongly coupled. In this case a one-to-one map from the space of real and symmetric tensors $h_{\mu\nu}$ to the space of metrics of fixed signatures indeed exists, since the perturbative nature of metric fluctuations $h_{\mu\nu}$ makes it difficult to flip the signature of the background metric $\bar{g}\equiv \eta$. Thus, no changes in the signature of  $g_{\mu\nu}=\eta_{\mu\nu}+h_{\mu\nu}$ are expected to occur.

A different approximation to evaluate the path integral~\eqref{eq:PIs} consists of integrating over symmetry-reduced configurations only. For example, in a minisuperspace setup the action is reduced to a function of just one variable. The quantization of the resulting action defines the minisuperspace approach to quantum gravity -- widely used in quantum cosmology \cite{Hawking:1983hn, Hawking:1983hj, Vilenkin:1994rn, Ashtekar:2011ni, Bojowald:2015iga} -- and allows to explore the impact of quantum gravity in the early evolution of the universe in a simplified setup  with minimal extra assumptions~\cite{Bojowald:2001xe, Ashtekar:2006rx, Hartle:2007gi, Feldbrugge:2017kzv, Feldbrugge:2017fcc, DiazDorronsoro:2017hti, DiazDorronsoro:2018wro, DiTucci:2019xcr, DiTucci:2019bui, Basile:2021euh, Basile:2021krk, Asante:2021phx}.
 
Finally, asymptotically safe quantum gravity aims at a fully non-perturbative evaluation of a path integral over metric configurations \cite{Weinberg:1980gg} based on diffeomorphism-invariant gravitational actions. Practical computations are typically performed either using \FRG{} \cite{Reuter:1996cp, Percacci:2017fkn, Reuter:2019byg, Dupuis:2020fhh} or lattice techniques \cite{Laiho:2017htj, Loll:2019rdj}.
In contrast to \EDT{} and \CDT{}, the \FRG{} is a continuum method implementing a local coarse-graining procedure. It is based on the scale-dependent effective action $\Gamma_k$, which interpolates between the classical action $S$ and the full quantum effective action $\Gamma_{k=0}=\Gamma_{\rm 1PI}$. The insertion of a regulator functional into the path integral implements the Wilsonian idea of momentum-shell-wise integration of quantum fluctuations. The definition of momenta requires the definition of a reference background metric $\bar{g}$. However, this background in principle never needs to be specified explicitly. In contrast to the \EFT{} treatment of quantum gravity, in asymptotically safe quantum gravity, metric fluctuations $h_{\mu\nu}$ can be of arbitrary size. It needs to be stressed at this point that, for technical reasons, the local coarse-graining procedure requires the use of Euclidean signature. The more challenging investigation of Lorentzian Asymptotic Safety with the \FRG{} is one of the open questions outlined in \cite{Bonanno:2020bil}. First steps towards a Lorentzian formulation have been taken in \cite{Manrique:2011jc, Rechenberger:2012dt, Biemans:2016rvp, Biemans:2017zca, Houthoff:2017oam, Knorr:2018fdu, Eichhorn:2019ybe, Bonanno:2021squ, Fehre:2021eob, Banerjee:2022xvi, DAngelo:2022vsh}.

\section{Parameterizations of metric fluctuations for Euclidean quantum gravity}\label{sec:Eucl-param}

In this section we discuss the Euclidean version of the path integral \eqref{eq:PIs}, assuming a suitable regularization. Our goal is to construct a bijective map between the full Euclidean metric $g_{\mu\nu}$ of a given topology, and  metric fluctuations $h_{\mu\nu}$ around a (fixed but arbitrary) Euclidean background $\bar{g}_{\mu\nu}$ of the same topology, such that the Euclidean path integral \eqref{eq:PIs} can be replaced by a regularized path integral over the metric fluctuations $h_{\mu\nu}$. The role of a fixed topology will be discussed below. This mapping can in principle be implemented using a linear parameterization, together with an inequality constraint, which only selects spacetimes of a given signature. However, if the latter is not properly implemented, for example due to approximations, the path integral over metric fluctuations can contain degenerate or Lorentzian metrics, cf.~subsection \ref{sec:exEuclid}. By contrast, in perturbative approaches to quantum gravity, for example in the context of an \EFT{} description of quantum gravity \cite{Donoghue:1993eb, Donoghue:1994dn, Bjerrum-Bohr:2002gqz, Donoghue:2015nba}, as well as quantum gravity in $2+\epsilon$ dimensions \cite{Tsao:1977tj, Duff:1977ay, Christensen:1978sc}, the linear parameterization without additional constraints is under control. As already mentioned, metric fluctuations are small in the regime of validity of the perturbative expansion and cannot change the signature of the background metric. We will discuss the possibility of employing an inequality constraint beyond the perturbative regime and its practical implementation in section \ref{sec:ineq-constraints}.

In this section we will show that in the Euclidean case there exist infinitely many bijective maps $f(\cdot)$ between the space of Euclidean metric $g_{\mu\nu}$ and the space of symmetric, real fluctuation fields $h_{\mu\nu}$, which takes the form
\begin{equation}\label{eq:genpara}
g_{\mu\nu} = \bar{g}_{\mu   \kappa}f(h^{\cdot}_{\phantom{\cdot}\cdot})^{\kappa}_{\,\,\nu}\,.
\end{equation}
Here indices on $h$ are raised and lowered with the background metric $\bar{g}$, and the full metric $g$ and the background metric $\bar{g}$ are defined on the same connected component, \ie{}, they share the same topology. In the following we will restrict ourselves to ultralocal parameterizations, \ie{}, those that do not include derivatives. Furthermore, in order to make a proper connection to a perturbative expansion, we will require that for small $h$
\begin{equation}\label{eq:f-pert}
f(h^\cdot_{\phantom{\cdot}\cdot})^\mu_{\phantom{\mu}\nu} = \delta^\mu_\nu + h^\mu_{\phantom{\mu}\nu} + \mathcal O(h^2) \, .
\end{equation}
In the next subsection we will study the conditions the parameterization $f$ has to satisfy in order to preserve the signature and to map bijectively between Euclidean metrics and real symmetric matrices. We will find that there are infinitely many parameterizations $f$ that satisfy these conditions.

Before we dive into the discussion, let us briefly clarify the role of the spacetime topology of the metric. As already mentioned, in this work we aim at finding maps relating metrics of the same, fixed topology. This is because with finite fluctuations, the topology cannot be ``changed''.\footnote{Here we refer to the change of topology of the four-dimensional spacetime metrics induced by their fluctuations $h$, and not to the more commonly discussed change of the spatial, three-dimensional topology as a function of time.} Let us stress that restricting the parameterization $f$ to be bijective ensures that $h$ is finite. In contrast, if one would allow $h$ to diverge, one would be able to connect metrics of different topologies, contradicting the notion that geometries having different topologies cannot be deformed continuously into each others. One can also illustrate this by considering a topological invariant, \eg{}, the Euler characteristic. Its variation with respect to the metric vanishes exactly, since it is a total derivative. This directly implies that finite quantum fluctuations cannot change the topology if we do not take into account boundary degrees of freedom.\footnote{Strictly speaking, this would still allow fluctuations which keep the Euler characteristic fixed but nevertheless change the topology. Here we make the assumption that topology changes are always tied to a change in an index that in general is contained in the action, and whose metric variation vanishes exactly.} An inclusion of such boundary terms would be very interesting and could potentially yield topology changes, but we leave this for future work.

\subsection{Existence of bijective, signature-preserving parameterizations for Eu\-cli\-de\-an gravity}

A necessary condition for bijectivity of the parameterization \eqref{eq:genpara} is to ensure that $g$ and $\bar{g}$ have the same signature. To investigate consequences of this condition, we use Sylvester's law of inertia \cite{Sylvester:1852}. It states that a symmetric square matrix $g_{\mu\nu}$ and another symmetric square matrix $\bar g_{\mu\nu}$ have the same number of positive, negative and zero eigenvalues if and only if they are related by
\begin{equation}\label{eq:SilvLaw}
 g = S^T \bar g S \, ,
\end{equation}
where $S$ is a real invertible matrix. At first glance, this appears to be a non-covariant relation. However, assuming that $S$ only depends covariantly on the matrix $\bar g^{-1} h$, we show in appendix \ref{sec:Squared} that $S$ has the form of a rank-(1,1) tensor, and that the relation \eqref{eq:SilvLaw} is equivalent to
\begin{equation}\label{eq:gdef}
 g_{\mu\nu} = \bar g_{\mu\rho} \left[ S(\bar g^{-1}h)^2 \right]^\rho_{\phantom{\rho}\nu} \, .
\end{equation}
This is the same as
\begin{equation}\label{eq:s-squared}
S^\mu_{\phantom{\mu}\rho} S^\rho_{\phantom{\rho}\nu}=\bar g^{\mu\rho} g_{\rho\nu} =f(h^\cdot_{\phantom{\cdot}\cdot})^\mu_{\phantom{\mu}\nu} \, ,
\end{equation}
where we suppressed the arguments, and where we have used \eqref{eq:genpara} for the last equality. This implies that parameterizations $f$ that preserve the signature have to be the square of an invertible matrix $S$. 

In what follows we discuss the conditions that bijectivity imposes on the parameterization $f$. Specifically, we will use the fact that positive-definite matrices like the Euclidean metrics $g$ and $\bar g$ possess a unique positive-definite square root.\footnote{Note that the root is, in general, not a rank-(0,2) tensor, since the matrix product of two rank-(0,2) tensors is not a tensor. This is in contrast to the root of a rank-(1,1) tensor, which can be a rank-(1,1) tensor again. For that reason, we will suppress the indices in the following. Since final expressions are covariant, all potential different choices for the branch of the square root lead to the same result.} Let us multiply \eqref{eq:s-squared} from the left by this unique square root of $\bar g$, and from the right by its inverse,
\begin{equation}
 \sqrt{\bar g^{-1}} \, g \, \sqrt{\bar g^{-1}} = \sqrt{\bar g} \, f(\bar g^{-1} \, h) \, \sqrt{\bar g^{-1}} \, .
\end{equation}
We now assume that $f$ and $S$ have the following property:
\begin{align}
	\sqrt{\bar g} f(\bar g^{-1}h) \sqrt{\bar g^{-1}} = f(\sqrt{\bar g^{-1}} h \sqrt{\bar g^{-1}}) \, , \label{eq:assumptSf} \\
	\sqrt{\bar g} S(\bar g^{-1}h) \sqrt{\bar g^{-1}} = S(\sqrt{\bar g^{-1}} h \sqrt{\bar g^{-1}}) \, . \label{eq:Sproperty}
\end{align}
This property is satisfied if $f$ and $S$ have a convergent Taylor expansion, or if they have a representation in terms of an integral transform with exponential kernel (\eg{}, a Laplace or Fourier transform).\footnote{Note that strictly speaking $f$ and $S$ also depend on the background metric $\bar g$ individually, since $f=\bar g$ for $h=0$. We will suppress this additional dependence in the following to avoid cluttered notation.}

Under the assumption that \eqref{eq:assumptSf} holds, it follows that
\begin{equation}
\label{eq:bijmapprelim}
 \sqrt{\bar g^{-1}} \, g \, \sqrt{\bar g^{-1}} = f(\sqrt{\bar g^{-1}} h \sqrt{\bar g^{-1}}) \equiv S(\sqrt{\bar g^{-1}} h \sqrt{\bar g^{-1}})^2 \, ,
\end{equation}
where the last equality follows from \eqref{eq:s-squared}.
Furthermore, we observe that the left-hand side of \eqref{eq:bijmapprelim} is a positive-definite matrix, since $g$ is positive-definite and for any vector $x$,
\begin{equation}\label{eq:x-pos-def}
 x^T \sqrt{\bar g^{-1}}\, g \,\sqrt{\bar g^{-1}} x = \left( \sqrt{\bar g^{-1}}\, x \right)^T g \left( \sqrt{\bar g^{-1}} \,x \right) > 0 \, .
\end{equation}
Hence, the argument of both $f$ and $S$ is a real symmetric matrix. We conclude that $f$ is a bijective map from real symmetric matrices to the space of positive-definite matrices. Therefore, we can also take its positive-definite square root to uniquely define $S$. This then ensures that $g$ and $\bar g$ have the same signature.

Since $f$ is bijective, we can solve \eqref{eq:bijmapprelim} for $h$ and get
\begin{equation}\label{eq:h_of_g}
 h = \sqrt{\bar g} \, f^{-1}(\sqrt{\bar g^{-1}} \, g \, \sqrt{\bar g^{-1}}) \, \sqrt{\bar g} \, .
\end{equation}
It is obvious that the $h$ defined by this equation is symmetric and real. It is also a tensor, which can be seen by expanding $f^{-1}$ in a power series. This exists at least locally due to the inverse function theorem and the perturbative constraint \eqref{eq:f-pert}, which allows us to write
\begin{equation}
 f^{-1}(\sqrt{\bar g^{-1}} \, g \, \sqrt{\bar g^{-1}}) = \sqrt{\bar g^{-1}} \, \left( g - \bar g \right) \sqrt{\bar g^{-1}} + \mathcal O ((g-\bar g)^2) \, ,
\end{equation}
so that
\begin{equation}
 h = g - \bar g + \mathcal O ((g-\bar g)^2) \, .
\end{equation}
This expansion moreover proves that the specific choice of square root is inessential, so that the relation between $g$ and $h$ is indeed unique. We can also write
\begin{equation}\label{eq:g_of_h}
 g = \sqrt{\bar g} \, f(\sqrt{\bar g^{-1}} h \sqrt{\bar g^{-1}}) \, \sqrt{\bar g} \, .
\end{equation}
This form makes it clear that $g$ is positive-definite, cf.~\eqref{eq:x-pos-def}, and indeed a Euclidean metric. With the condition stated above that $f$ is a bijection between real symmetric matrices and positive-definite matrices, equations \eqref{eq:h_of_g} and \eqref{eq:g_of_h} conclude the discussion on bijectivity.

\subsection{Examples, discussion and implications}
\label{sec:exEuclid}

In this section we explicitly show that the linear parameterization violates the previously derived conditions. Furthermore, we derive an entire family of bijective parameterizations and discuss the importance of using a bijective parameterization to avoid inequality constraints. Finally, we discuss some caveats on the practical implementation of our construction.

Summarizing the findings of the previous subsection, requiring a bijective map between Euclidean metrics $g$ and real symmetric tensors $h$ restricts the allowed parameterizations $f$, defined in \eqref{eq:genpara}, in the following way:
\begin{itemize}
 \item $f$ maps real symmetric matrices bijectively to positive-definite real and symmetric matrices, and
 \item for small arguments,
  \begin{equation}
   f(\bar g^{-1} h) = \mathbbm 1 + \bar g^{-1} h + \mathcal O(h^2) \, ,
  \end{equation}
  to abide with the perturbative expansion, \eqref{eq:f-pert}.
\end{itemize}
The first condition, together with \eqref{eq:Sproperty}, implies that $S$, the unique positive-definite square root of $f$, is invertible as a matrix, such that via Sylvester's law of inertia, $\bar{g}$ and $g$ have the same signature under such a parameterization.

Let us discuss some concrete examples. First, the linear parameterization 
\begin{equation}
 g_{\mu\nu} = \bar{g}_{\mu\nu} + h_{\mu\nu} \, ,
\end{equation}
fails the first criterion, since
for the particular matrix $h=-\bar g$, the metric $g$ vanishes. 
Second, as has been established before \cite{Nink:2014yya}, the exponential parameterization
\begin{equation}
\label{eq:expparam}
 f^{\rm{exp}}(\bar g^{-1} h) = \exp \left( \bar g^{-1} h \right) \, 
\end{equation}
fulfills all requirements.

One can construct an entire family of viable parameterizations inspired by the exponential one. In particular,
\begin{equation}\label{eq:b-expparam}
 f^{\rm{exp-b}}(\bar g^{-1} h) = \exp \left( \bar g^{-1} h^{\rm{TL}} \right) b(h^{\rm{Tr}}) \, 
\end{equation}
is a family of viable bijections whenever $b$ is a bijection from $\mathbbm R$ to $\mathbbm R_+$, where we have split the fluctuation $h$ into traceless ($h^{\rm{TL}}$) and trace ($h^{\rm{Tr}}$) parts,
\begin{equation}
 h = h^{\rm{TL}} + \frac{1}{d} \bar g \, h^{\rm{Tr}} \, , \qquad {\rm{tr}} \left( \bar g^{-1} h^{\rm{TL}} \right) = 0 \, .
\end{equation}
Herein, $d$ is the dimension of the Euclidean spacetime.
To show that this indeed provides a viable parameterization and to derive the form of $b$, we insert the ansatz given by \eqref{eq:b-expparam} into the definition of the parameterization \eqref{eq:genpara}, which gives
\begin{equation}
\label{eq:famparam}
 g_{\mu\nu} = \bar g_{\mu\rho} \exp \left( \bar g^{-1} h^{\rm{TL}} \right)^\rho_{\phantom{\rho}\nu} \, b(h^{\rm{Tr}}) \, .
\end{equation}
Multiplying by $\bar g^{-1}$ from the left and taking the determinant, we get
\begin{equation}
 \det \bar g^{-1} g = b(h^{\rm{Tr}})^d \, .
\end{equation}
Since by assumption both $\bar{g}$ and $g$ are positive-definite, the left-hand side is positive and we can take the unique real  positive $d$-th root, which determines $b(h^{\rm{Tr}})$. Using this, we can also solve for the traceless component,
\begin{equation}
 \exp \left( \bar g^{-1} h^{\rm{TL}} \right) = \left( \det \bar g^{-1} g \right)^{-1/d} \bar g^{-1} g \, .
\end{equation}
The proof of bijectivity then follows as above.

Let us briefly address the question of practical importance for studies of the quantum gravitational path integral. Our starting point is the transition from a path integral over the metric $g$ to a path integral over metric fluctuations $h$ on a background metric $\bar{g}$. We have argued that choosing a parameterization fulfilling the above conditions is necessary to avoid an inequality constraint in the path integral. This inequality constraint would, for an arbitrary parameterization, fix the signature of the metric and avoid overcounting metric configurations. In general, the direct implementation of such an inequality constraint would give rise to a Heaviside distribution in the path integral measure. Besides the difficulties of the practical implementation (\eg{}, making sure that the inequality constraint is implemented exactly even in settings which require approximations), it is in general non-trivial to provide the explicit form of the constraint. In addition to the inequality constraint, one must make sure that no physical configuration is left out by any parameterization, be it of the above form or not.  

As we will show in section \ref{sec:Lorentzian}, signature-preserving parameterizations of the metric that can be straightforwardly used for Euclidean signature fail for Lorentzian spacetimes. One therefore must either use inequality constraints (or other more exotic techniques) or accept a path integral which takes both Euclidean and Lorentzian metrics into account. We will come back to this construction in section \ref{sec:Lorentzian}, with the concrete implementation of the inequality constraint in section \ref{sec:ineq-constraints}.

Finally, parameterizations as those discussed above only provide a bijective mapping if fluctuations of arbitrary order are taken into account (assuming that $f$ is not a polynomial). For instance, if one only works to first order in $h$ in the exponential parameterization, one recovers the linear parameterization, such that the signature preservation is no longer guaranteed. More generally, once the series expansion for a given $f$ is truncated at some finite order, a finite $h_{\mu\nu}$ can, \eg{}, still flip the signature. On a practical level however, the inclusion of higher and higher powers will eventually converge to the true bijective parameterization, whereas for, \eg{},  the linear parameterization, we expect at best a finite radius of convergence in $h$. Indeed,  such results were found in explicit studies:
A non-perturbative renormalization group study of quantum gravity that included an arbitrary power of metric fluctuations in the exponential parameterization \cite{Knorr:2017mhu} found an infinite radius of convergence. An analogous calculation in the linear parameterization reveals a singularity in the \RG{} equation at $h=-\bar g$, as expected from the discussion above.

\section{Constraints for foliated spacetimes} \label{sec:foliation}

In a path integral formulation of quantum gravity, globally hyperbolic spacetimes play a special role, since they possess a well-defined causal structure. This causal structure ensures the existence of a well-defined Wick rotation to Euclidean spacetimes. Therefore, if the Lorentzian path integral is restricted to a sum over globally hyperbolic spacetimes, only these configurations contribute to physical scattering amplitudes. This is indeed the philosophy at the basis of the \CDT{} program \cite{Loll:2000my}.

In the context of \CDT{}, the foliation structure is implemented by means of the Arnowitt-Deser-Misner (\ADM{}) formalism \cite{Arnowitt:1962hi}, as this guarantees that only ``foliatable'' spacetimes are included in the sum over paths. A similar implementation in the context of the \FRG{} approach to asymptotically safe quantum gravity \cite{Rechenberger:2012dt, Biemans:2016rvp, Platania:2017djo, Houthoff:2017oam} would however either break diffeomorphism invariance down to foliation-preserving diffeomorphism \cite{Rechenberger:2012dt}, or break the one-loop structure of the \FRG{} equation \cite{Knorr:2018fdu}.

Motivated by these considerations, in this section we will discuss the implementation of a foliation structure in a manifestly covariant way and within the background field formalism, extending previous work \cite{Knorr:2018fdu} to general signature-preserving parameterizations. We will review some basic properties of the foliation structure and the split of the full fields into background and fluctuations. In this analysis, we will focus on \emph{Euclidean} signature, using the results of the previous section. Our goal is to formulate a path integral over foliatable manifolds in terms of $h_{\mu\nu}$ only.

There are two important aspects to be kept in mind when interpreting the results in the following in a broader context. First, demanding the existence of a foliation in Euclidean signature is different from requiring  the existence of a consistent causal structure. As a matter of fact, a foliation in Euclidean signature is always associated to a spatial direction, but not necessarily to the spatial direction along which an analytical continuation to Minkowski space is performed. Therefore, restricting the configuration space in Euclidean signature to foliatable configurations only does not automatically result in the configuration space of Wick-rotated Lorentzian spacetimes (assuming that an analytical continuation can be performed at all, see \cite{Baldazzi:2018mtl}), unless additional conditions are suitably implemented. Second, since \textit{locally} any Riemannian manifold is ${\mathbb{R}}^4$, one can always foliate it locally. Therefore, any difference between the configuration spaces of foliatable and non-foliatable Riemannian manifolds must be a \emph{global} one.

\subsection{Basics}\label{sec:params}

In this section we lay out the basics to implement a foliation on a Euclidean spacetime in a covariant way, and discuss fluctuations in terms of ``foliation variables'' -- a vector $n_\mu$, defining the foliation, and the induced metric on the spatial slices, $\sigma_{\mu\nu}$. Furthermore, we discuss the importance of a linear relation between the metric fluctuations $h_{\mu\nu}$ and the fluctuations of the foliation variables in the context of the \FRG{}. We will come to the conclusion that, in Euclidean signature, the existence of a foliation does not impact the path integral. 

We start by assuming that there exists a foliation of the spacetime metric $g_{\mu\nu}$ in terms of the orthogonal normalized vector $n_{\mu}$ and the spatial metric $\sigma_{\mu\nu}$ according to \cite{Knorr:2018fdu}
\begin{equation}\label{eq:metricfoliation}
g_{\mu\nu} = \sigma_{\mu\nu} + n_{\mu} n_{\nu} \, ,
\end{equation}
with the conditions that
\begin{equation}
\label{eq:conds}
n_{\mu} \, g^{\mu\nu} \, \sigma_{\nu\rho} = 0 \, , \qquad n_{\mu} \, g^{\mu\nu} \, n_{\nu} = 1 \, .
\end{equation}
The first equation states that $n_{\mu}$ and $\sigma_{\mu\nu}$ are orthogonal, whereas the second equation provides a normalization for $n_{\mu}$. 
We stress that in this covariant framework, both $n_\mu$ and $\sigma_{\mu\nu}$ are four-dimensional covariant objects, which distinguishes this setup from the \ADM{}-framework \cite{Arnowitt:1960es, Arnowitt:1962hi}, where the spatial metric and the shift vector are typically introduced with spatial indices only. The decomposition of the full metric $g$ in terms of foliation variables $\sigma$ and $n$ according to \eqref{eq:metricfoliation} increases the number of dynamical degrees of freedom. The role of the two conditions \eqref{eq:conds} is to ensure that the same number of dynamical degrees of freedom exists. We will comment on this point more extensively in section \ref{sec:ConstFol}.

We split the metric into a fixed but arbitrary background $\bar{g}_{\mu\nu}$ and a fluctuation 
$h_{\mu\nu}$. The corresponding fluctuations of $n_{\mu}$ will be called $\hat{n}_{\mu}$, those of $\sigma_{\mu\nu}$ will be called $\hat{\sigma}_{\mu\nu}$. Within a background field formalism, $\hat{n}$ and $\hat{\sigma}$ are the integration variables in the path integral. Given that $n$ and $\sigma$ are not independent, one cannot simply assume a linear relationship between $n$ and $\hat{n}$, or $\sigma$ and $\hat{\sigma}$.
In fact, the dependence of the full fields on the fluctuations can be non-linear, and even mixed, so that we write
\begin{equation}\label{eq:decomposition}
n_{\mu} = \bar{n}_{\mu} + \delta n_{\mu}(\hat{n},\hat{\sigma}) \, , \qquad \sigma_{\mu\nu} = \bar{\sigma}_{\mu\nu} + \delta\sigma_{\mu\nu}(\hat{n},\hat{\sigma}) \, ,
\end{equation}
where $\delta n_{\mu}$ and $\delta\sigma_{\mu\nu}$ are the parameterizations of foliation fluctuations. They play the same role as the parameterization $f$ introduced in the previous section does for the metric fluctuation. 
We will assume in the following that the background quantities satisfy the background version of \eqref{eq:conds}. 

With the decomposition \eqref{eq:decomposition}, the full metric as a function of parameterized foliation fluctuations reads
\begin{equation}\label{eq:metricfoliationBGplusFLUC}
g_{\mu\nu} = \bar{g}_{\mu\nu} + \delta\sigma_{\mu\nu} + \bar{n}_{\mu} \delta n_{\nu} + \delta n_{\mu}  \bar{n}_{\nu} + \delta n_{\mu} \delta n_{\nu} \, .
\end{equation}
We now want to relate the foliation fluctuations $\hat{n}$ and $\hat{\sigma}$ to the metric fluctuation $h$ in a way that, at variance with the \ADM{} formalism, preserves general covariance explicitly.

Demanding a linear relationship between these two types of fluctuations, the unique relation between metric fluctuations $h$ and foliation fluctuations $\hat{n}$ and $\hat{\sigma}$ is
\begin{equation}\label{eq:htofolfluc}
h_{\mu\nu} = \hat{\sigma}_{\mu\nu} + \bar{n}_{\mu} \hat{n}_{\nu} + \hat{n}_{\mu} \bar{n}_{\nu} \, .
\end{equation}
Let us stress that in a general setting, a linear relation might not be strictly necessary. One specific framework where the linearity is critical is the \FRG{}. In this context, the path integral is regularized by means of a mode-suppression term that is quadratic in the field fluctuations. It thus acts as a scale-dependent mass term $h\cdot R(p^2/k^2)\cdot h$, where $k$ is a \RG{} scale. Starting from a path integral where the gravitational action is written and regularized in terms of $h$, and performing the transformation to foliation variables in a second step, only a linear relation can preserve the quadratic nature of the mode-suppression term. This quadratic structure is key in order for the exact \FRG{} equation derived in \cite{Wetterich:1992yh, Morris:1993qb, Ellwanger:1993mw}, and adapted for gravity in \cite{Reuter:1996cp}, to be of one-loop form.

Alternatively, one might take the point of view that irrespective of the relationship between $h$ and $(\hat{n},\hat{\sigma})$, one can simply introduce a mode-suppression term that is quadratic in the foliation fluctuations $\hat{n}$ and $\hat{\sigma}$. Yet, this comes at the cost of broken background diffeomorphism invariance, unless the relationship between $h$ and $\hat{n}$, $\hat{\sigma}$ is linear, and the regulators of the individual sectors are tuned in the right way. This is because both the background metric $\bar{g}$ as well as the fluctuation field $h$ transform under a background diffeomorphism, such that this auxiliary symmetry becomes the full gauge symmetry once $\bar{g}$ and $h$ are combined to the metric $g$ again. Therefore, $\hat{n}$ and $\hat{\sigma}$ transform under background diffeomorphisms. Unless the relation between $h$ and $\hat{n}$, $\hat{\sigma}$ is linear, they transform in a non-linear fashion under the auxiliary symmetry. Accordingly, a regulator that is quadratic in these fields and is introduced in an ad hoc manner in the case of a non-linearly realized symmetry, breaks background diffeomorphism symmetry. The latter is critical to recover a diffeomorphism invariant effective action from the \FRG{} formalism. Therefore, the linearity of the relation \eqref{eq:htofolfluc} is critical in this setup.

\subsection{Constraints on foliation fluctuations}
\label{sec:ConstFol}

In this section we investigate the implications of the constraints \eqref{eq:conds}, which ensure orthogonality of $n$ and $\sigma$, as well as the normalization of $n$, on the foliation fluctuations. Demanding that the parameterization of fluctuations of the normal vector $\delta n$ is completely determined by the metric fluctuation $h$, there are parameterizations of foliation variables that satisfy the foliation constraints \eqref{eq:conds} and at the same provide a bijective map between background and full metric.

We start with the second constraint, which reads 
\begin{equation}
\mathcal{F}_2=n_{\mu} \,g^{\mu\nu} \, \sigma_{\nu\rho} = 0 \, .
\end{equation}
Using the decomposition \eqref{eq:metricfoliation}, we see that the two constraints  \eqref{eq:conds} are not independent,
\begin{equation}
\mathcal F_2 = n_{\mu} \,g^{\mu\nu} \, \sigma_{\nu\rho} =  n_{\mu} \,g^{\mu\nu} \, (g_{\nu\rho} - n_{\nu} n_{\rho}) = \left( 1 -  n_{\mu} \,g^{\mu\nu} \, n_{\nu} \right) n_{\rho} = -\mathcal F_1 n_{\rho} \, .
\end{equation}
In particular, it is just a single scalar constraint that is effectively imposed.
The orthogonality and normalization conditions \eqref{eq:conds} are thus not enough to remove all unphysical fluctuations. By counting the modes, the set of foliation fluctuations $(\hat{n}_{\mu}, \hat{\sigma}_{\mu\nu})$ has 14 independent components, while the metric fluctuation $h_{\mu\nu}$ only carries 10 (before gauge fixing). It follows that the constraints have to remove four unphysical modes that have been introduced by the decomposition \eqref{eq:metricfoliation}. However, as we have seen, the vectorial constraint and the scalar constraint collapse to one single scalar constraint. This is in fact a consequence of the invariance of the metric $g_{\mu\nu}$ in \eqref{eq:metricfoliationBGplusFLUC} under the shift of the parameterization of foliation fluctuations $\delta n_\mu$ and $\delta \sigma_{\mu\nu}$ by some vector field 
$\omega_{\mu}$,
\begin{equation}\label{eq:Z-shift-invariance}
\delta n_{\mu} \to \delta n_{\mu} + 
\omega_{\mu} \, , \qquad \delta\sigma_{\mu\nu} \to \delta\sigma_{\mu\nu} - (\bar{n}_{\mu} + \delta n_{\mu})
\omega_{\nu} -
\omega_{\mu} (\bar{n}_{\nu} + \delta n_{\nu}) - 
\omega_{\mu} 
\omega_{\nu} \, .
\end{equation}
This corresponds to a change of the foliation at fixed total metric $g$.
Accordingly, it is enough to satisfy the scalar normalization constraint in order to implement a foliation structure. 

At this point, we are left with the task of removing the three remaining unphysical modes introduced by the foliation variables. To this end, it is actually preferable to replace the fluctuation of the spatial metric by the metric fluctuation $h$ via \eqref{eq:metricfoliationBGplusFLUC}. Using $h$ has the virtue that for a diffeomorphism invariant theory, all information is carried by $h$ only, while $n$ cannot appear in any equation. In fact, $n$ can only appear explicitly in the dynamics if the symmetry is explicitly broken to foliation-preserving diffeomorphisms.

To fix the mismatch in the number of fluctuating fields, we require that the parameterization of fluctuations of the normal vector $n$ is completely determined by the metric fluctuation (and background quantities),\footnote{This requirement is not unique, and one could replace it by other conditions to fix the mismatch in the number of degrees of freedom. We employ this choice as the simplest and most convenient one.}
\begin{equation}\label{eq:Soln}
\delta n_{\mu} = \bar{n}_{\rho} \, \mathbbm X(h^{\,\cdot}_{\phantom{\,\cdot}\cdot})^{\rho}_{\phantom{\rho}\mu} \equiv \bar{g}_{\mu\rho} \, \mathbbm X(h^{\,\cdot}_{\phantom{\,\cdot}\cdot})^{\rho}_{\phantom{\rho}\kappa} \, \bar{g}^{\kappa\alpha} \, \bar{n}_{\alpha} \, ,
\end{equation}
and we fix the matrix $\mathbbm X$ by requiring that the constraint
\begin{equation}
\label{eq:cons}
n_{\mu} \, g^{\mu\nu} \, n_{\nu}=\bar{n}_{\mu} \, \bar{g}^{\mu\nu} \,\bar{n}_{\nu}=1 \, ,
\end{equation}
is satisfied. Inserting the ansatz for the parameterization of fluctuations of the normal vector $n$, \eqref{eq:Soln}, on the left-hand side yields
\begin{equation}
\label{eq:consins}
n_{\mu} \, g^{\mu\nu} \, n_{\nu}=\bar{n}_{\mu}\left(\delta^{\mu}_{\phantom{\mu}\alpha}+\mathbbm X^{\mu}_{\phantom{\mu}\alpha}\right)\,g^{\alpha\rho}\,\bar{g}_{\rho\kappa}\,\left(\delta^{\kappa}_{\phantom{\kappa}\lambda}+\mathbbm X^{\kappa}_{\phantom{\kappa}\lambda}\right)\bar{g}^{\lambda\sigma}\,\bar{n}_{\sigma} \, ,
\end{equation}
from which we conclude by comparison with the right-hand side that
\begin{equation}
\begin{aligned}
 &\delta^{\mu}_{\phantom{\mu}\lambda}= \left(\delta^{\mu}_{\phantom{\mu}\alpha}+\mathbbm X^{\mu}_{\phantom{\mu}\alpha}\right)\,g^{\alpha\rho}\,\bar{g}_{\rho\kappa}\,\left(\delta^{\kappa}_{\phantom{\kappa}\lambda}+\mathbbm X^{\kappa}_{\phantom{\kappa}\lambda}\right) \\
 & \Rightarrow \left(\delta^{\mu}_{\phantom{\mu}\rho}+\mathbbm X^{\mu}_{\phantom{\mu}\rho}\right)\left(\delta^{\rho}_{\phantom{\rho}\nu}+\mathbbm X^{\rho}_{\phantom{\rho}\nu}\right) = \bar{g}^{\mu\kappa}\, g_{\kappa\nu} \equiv S^\mu_{\phantom{\mu}\kappa} S^\kappa_{\phantom{\kappa}\nu} \, .
\end{aligned}
\end{equation}
In the last step, we have used the condition that $\bar g$ and $g$ have the same signature as discussed in the previous section, cf.~\eqref{eq:s-squared}. We can now immediately fix the parameterization of fluctuations of the normal vector in terms of the parameterization of metric fluctuations:
\begin{equation}
 \mathbbm X = S - \mathbbm 1 \, .
\end{equation}
In this way, the tensor $\sigma_{\mu\nu}$ and the vector $n_{\mu}$ satisfy the constraints \eqref{eq:conds}, and at the same time the parameterization of foliation fluctuations $\delta n_{\mu}$ and $\delta \sigma_{\mu\nu}$ are expressed entirely in terms of background quantities and $h_{\mu\nu}$. This means that the metric fluctuations $h_{\mu\nu}$ remain completely unconstrained, so that no inequality constraints on $h_{\mu\nu}$ are necessary on the level of the path integral.

\subsection{Discussion and interpretation}

As discussed in section \ref{sec:Eucl-param}, the use of suitable parameterizations for the metrics fluctuations $h$ which preserve the signature of the full metric $g$ makes it possible to avoid a direct implementation of an inequality constraint. Again, this property is automatically ensured by using a map satisfying \eqref{eq:SilvLaw}, as this leaves the signature of the full metric, and therefore also the signature of the spatial metric, unchanged. Examples of these maps are the exponential parameterization, \eqref{eq:expparam}, and the family of maps in \eqref{eq:b-expparam}.

We might therefore interpret the results of this section as an indication that the use of a suitable parameterization for the metric fluctuations combined with the choice of a foliatable background already implement the foliation structure for the full metric. 

We emphasize that the conditions we have derived only ensure the existence of some foliation for the full metric. 
This does not ensure that some appropriate analytical continuation of a set of Lorentzian spacetimes is included. It only ensures that there is a ``time direction'' that can be identified in the Euclidean setting, such that an analytical continuation would result in a Lorentzian manifold. Additionally, a Wick rotation is not straightforwardly implemented in quantum field theories on a curved spacetime,  see \cite{Baldazzi:2018mtl}.

\section{Parameterizations of metric fluctuations for Lorentzian quantum gravity} \label{sec:Lorentzian}

We have shown that it is possible to define the path integral over all foliatable Euclidean metrics in the background field formalism without introducing inequality constraints. This allows us to write a well-defined path integral which contains both a gauge fixing as well as a local regularization term. The setup enables studies of Euclidean Asymptotic Safety in a foliatable, covariant setting, where full background diffeomorphism symmetry is preserved. 

The real Universe is Lorentzian. It is thus key to understand whether it is possible to use the background field method in Lorentzian quantum gravity without the need for inequality constraints. In fact, in the case of quantum gravity (or even ``standard'' quantum field theory on a curved background) it is not clear whether it is always possible to ``connect''  Euclidean and Lorentzian setups via an analytic continuation. In particular, there are metrics which exist in Euclidean, but do not exist in Lorentzian signature, and vice versa. Additionally, even for simple examples like a de Sitter space, the analytical continuation is much more subtle than one might naively expect~\cite{Baldazzi:2018mtl}. It is therefore crucial to explore the case of Lorentzian quantum gravity directly.

In this section we will show that in the Lorentzian case, no ``simple'' bijective parameterizations can be found, and in particular, all parameterizations that work in the Euclidean case cannot work in the Lorentzian case. We cannot prove however that no bijective parameterizations exist -- their construction is nevertheless much more involved than in Euclidean signature.

Our starting point is the same as for Euclidean signature, namely that the (Lorentzian) metrics $g$ and $\bar g$ have the same signature if and only if there is a real invertible matrix $S$ such that
\begin{equation}\label{eq:SgSlor}
 g = S^T \, \bar g \, S \, .
\end{equation}
We will prove that for Lorentzian signature, $S$ in general cannot be a tensor. The proof proceeds by contradiction, thus assume that $S$ is a rank-(1,1) tensor. This is the only index structure such that the above equation makes sense. In appendix \ref{sec:Squared} we showed that if $S$ is a tensor, we can write the equivalent condition\footnote{More specifically, we have shown this under the assumption that $S$ only depends on $\bar{g}^{-1}h$ and the identity. In  principle, $S$ could depend on non-covariant objects that combine in such way that $S$ in the end is a tensor. We will not comment on this possibility and instead work under the reasonable assumptions that $S$ depends on $\bar{g}^{-1}h$ and the identity only.}
\begin{equation}\label{eq:Ssqlor}
 \bar g^{\mu\alpha} \, g_{\alpha\nu} = S^\mu_{\phantom{\mu}\alpha} S^\alpha_{\phantom{\alpha}\nu} \, .
\end{equation}
This implies that if $S$ is a tensor, then the combination $\bar g^{-1} \, g$ has to admit at least one real invertible square root. We will now construct an example where a specific choice of metrics does \emph{not} give rise to such a root. For this, consider the Kerr-Newman black hole of mass $M$ in ingoing Kerr coordinates,
\begin{equation}
\label{eq:Kerrmetricnew}
\begin{aligned}
ds^2 &= -\frac{ (\Delta-a^2 \sin^2\theta)}{\Sigma}dv^2 + 2dv dr-2 \frac{a \sin^2\theta}{\Sigma}\left(r^2+a^2-\Delta \right)dv d\chi \\
& \qquad - 2 a \sin^2 \theta d\chi dr + \frac{(r^2+a^2)^2- \Delta a^2 \sin^2\theta}{\Sigma}\sin^2\theta d\chi^2+\Sigma d\theta^2\,,
\end{aligned}
\end{equation}
where $\Delta = r^2 - 2Mr+a^2+Q^2$ and $\Sigma = r^2+a^2 \cos^2\theta$, $a$ denotes the spin and $Q$ the charge of the black hole. In the choice of the specific parameters $(M,Q,a)$ for the full metric $g$ and for the background metric $\bar{g}$, one has to make sure that both belong to the same topological class, since otherwise they might not be connected by a continuous parameterization $f(h)$. Since all four-dimensional non-extremal black holes are characterized by an Euler characteristic $\chi=2$~\cite{Wang:1998kb, PhysRevD.58.124026, PhysRevD.67.024027, Howard:2013yqq},
it suffices to choose $(M,Q,a)$ such that both $g$ and $\bar{g}$ have two non-degenerate horizons. Specifically, choosing the background metric $\bar g$ with parameters
\begin{equation}
 \bar M = 1 \, , \qquad \bar Q = \frac{3}{4} \, , \qquad \bar a = -\frac{1}{2} \, , 
\end{equation}
and the full metric $g$ with parameters
\begin{equation}
 M = 1 \, , \qquad Q = 0 \, , \qquad a = \frac{1}{2} \, ,
\end{equation}
and considering the point
\begin{equation}
 r = \frac{1}{10} \, , \qquad \theta = \frac{\pi}{2} \, ,
\end{equation}
which in both cases is located between the two horizons, we get for the product
\begin{equation}
 \bar g^{-1} \, g = \left( \begin{array}{rrrr} 1001&50&-526&0 \cr \frac{4225}{4}&51&-\frac{4433}{8}&0 \cr -2000&-100&1051&0 \cr 0&0&0&1 \end{array} \right) \, .
\end{equation}
Since the eigenvalues of this matrix are all distinct, it has exactly 16 distinct square roots. It is straightforward to show that none of the roots are real.

Nevertheless, since the two matrices have the same signature, there must be at least one real $S$ such that \eqref{eq:SgSlor} is fulfilled. The conclusion that we have to draw from the above observation is that \eqref{eq:Ssqlor} is not equivalent to this condition. This means that the $S$ relating Lorentzian metrics in general cannot be a tensor. From this statement we can draw an immediate conclusion: any parameterization that works in the Euclidean, where we have shown that $S$ is a tensor, cannot work in the Lorentzian.

More generally, one finds that those combinations of metrics $g$ and $\bar g$ do not admit a tensorial $S$ whose product $\bar g^{-1} g$ admits distinct negative eigenvalues. To see this, first note that, since both metrics have the same signature by assumption,
\begin{equation}
 \det \bar g^{-1} g > 0 \, .
\end{equation}
This implies that in four dimensions, the eigenvalues of this product can have the following form:
\begin{itemize}
	\item four positive eigenvalues,
	\item two positive, and two negative eigenvalues,
	\item four negative eigenvalues,
	\item two positive, and one pair of complex conjugate eigenvalues,
	\item two negative, and one pair of complex conjugate eigenvalues,
	\item two pairs of complex conjugate eigenvalues.
\end{itemize}
Now assume that we can diagonalize the product over the reals, so that
\begin{equation}
 P^{-1} \bar g^{-1} g P = P^{-1} S^2 P = (P^{-1} S P)^2 \, ,
\end{equation}
are both diagonal matrices. Now it is clear that if there are distinct negative eigenvalues, taking the square root will yield complex entries, so that $S$ isn't real. The qualifier for distinct negative eigenvalues is necessary since, \eg{}, the matrix
\begin{equation}
 \left( \begin{array}{rr} 0 & -1 \\ 1 & 0 \end{array} \right) \, ,
\end{equation}
is a real square root of minus the identity.

From this discussion we conclude that avoiding an inequality constraint by a suitable choice of parameterization seems to be very involved in the Lorentzian case. For that reason, in the next section we discuss the direct implementation of an inequality constraint in the context of the Wetterich equation.

We note that even restricting the metric to be globally hyperbolic likely does not solve the issue. As an example, we decompose the full metric and the (inverse) background metric as 
\begin{equation}
g_{\mu\nu} = \sigma_{\mu\nu} + n_\mu n_\nu \, , \qquad \bar g^{\mu\nu} = \bar \sigma^{\mu\nu} + n^\mu n^\nu \, ,
\end{equation}
such that the fluctuations of the normal vector $n$ vanish, which implies that $n=\bar{n}$. For these choices for $\bar{g}$ and $g$, $1$ is an eigenvalue of $\bar g^{-1} g$ with eigenvector $n$. Because of the orthogonality of $n$ to both $\bar{\sigma}$ and $\sigma$, the remaining eigenspace of $\bar g^{-1} g$ is spanned by spatial vectors, and we have to consider the eigenvalues of the product of two spatial Euclidean metrics $\bar{\sigma}^{-1}\sigma$.  Since $\bar{\sigma}^{-1}$ and $\sigma$ do not commute in general, this product is generally not symmetric. Therefore, it is not guaranteed that $\bar{\sigma}^{-1}\sigma$ features only positive eigenvalues. We conclude that even restricting the path integral to globally hyperbolic Lorentzian spacetimes only, a bijective, signature preserving parameterization of metric fluctuations of the form \eqref{eq:genpara} and \eqref{eq:SgSlor} does not exist.

\section{Inequality constraints in non-perturbative RG flows} \label{sec:ineq-constraints}

We have seen that in the Lorentzian case it is very difficult to find a parameterization for metric fluctuations which preserves the (Lorentzian) signature of the metric. In other words, it is difficult to restrict the gravitational path integral to Lorentzian spacetimes only. There are therefore three possibilities: i) in a quantum theory of gravity, fluctuations of the metric signature should be allowed, ii) the gravitational path integral must be restricted by means of inequality constraints or iii) we have to work within an altogether different configuration space which ensures classical equivalence with general relativity but where quantum fluctuations take a different form. In this section we discuss option ii), the implementation of inequality constraints. We will work within a toy model, and illustrate how an inequality constraint can be implemented directly within the \FRG{}. Extending this procedure beyond the simple toy model and using the linear parameterization, it might provide a suitable way to restrict the path integral to include Lorentzian signatures only.

In the literature, inequality constraints have often been avoided due to the difficulties arising in their implementation. As a first attempt to study quantum field theories subject to inequality constraints, in this section we focus on a simple scalar theory and we propose a method to implement an inequality constraint at the level of the \FRG{} equations. This method could be relevant for future studies of asymptotically safe quantum gravity in Lorentzian signature.
Let us  consider a scalar field $\phi$ whose maximum value is bounded from above, $\phi^2 \leq \phi_0^2$. Formally, this condition can be implemented by introducing a Heaviside distribution $\theta$ at the level of the Lorentzian path integral,
\begin{equation}
 \mathcal Z = \int \mathcal D \phi \, \theta\left(1 - \frac{\phi^2}{\phi_0^2} \right) e^{\mathbf{i} S[\phi]} \, ,
\end{equation}
where $S$ is the action of the scalar field. We will interpret the constraint as an addition to the action by exponentiating it. To avoid problems due to the non-differentiable nature of the Heaviside distribution, we first rewrite it as the limit of a smooth function,
\begin{equation}\label{eq:thetarep}
 \theta(x) = \lim_{\alpha\,\searrow\,0} \frac{1}{1+e^{-\frac{2x}{\alpha}}} \, .
\end{equation}
In this way,
\begin{equation}
 \mathcal Z = \int \mathcal D \phi \, \lim_{\alpha\,\searrow\,0} \exp \left[ \mathbf{i} S[\phi] - \ln \left( 1 + e^{-\frac{2}{\alpha}\left(1 - \frac{\phi^2}{\phi_0^2}\right)} \right) \right] \, ,
\end{equation}
such that the inequality constraint adds an additional imaginary component to the scalar potential. For a Euclidean path integral, the extra term is real.
We can now perform the standard derivation of the Wetterich equation (see, \eg, \cite{Gies:2006wv} for a detailed review of the derivation in the case of unconstrained systems, and \cite{Manrique:2011jc, Floerchinger:2011sc, Rechenberger:2012dt, Fehre:2021eob} for considerations on the Lorentzian case) to arrive at
\begin{equation}
 \partial_t \Gamma_k = \lim_{\alpha\,\searrow\,0} \frac{1}{2} {\rm{STr}} \left[ \frac{1}{\Gamma_k^{(2)} + R_k + \mathbf{i} \partial_\phi^2 \ln \left( 1 + e^{-\frac{2}{\alpha}\left(1 - \frac{\phi^2}{\phi_0^2}\right)} \right)} \partial_t R_k \right] \, .
\end{equation}
At the level of the flow equation, and after taking the second derivative, we can  take the limit  $\alpha\to0$. We have to discuss three different cases. For $\phi^2>\phi_0^2$, the second derivative of the constraint diverges like $1/\alpha$, and  therefore the contribution of these field configurations to the \RG{} flow is suppressed, as required by the inequality constraint. For $\phi^2<\phi_0^2$, the contribution of the constraint vanishes exponentially quickly, and thus the \RG{} flow is unaltered in the strict limit $\alpha\to0$. Finally, at the boundary $\phi^2 = \phi_0^2$, the constraint gives a contribution which diverges quadratically, as $\propto 1/\alpha^2$.\footnote{Let us note that the precise way in which the flow is suppressed at small but finite smearing parameter $\alpha$ depends on the specific function chosen to define the Heaviside distribution (in our case \eqref{eq:thetarep}), but the flow of $\Gamma_k$ obtained by taking the strict limit $\alpha\to0$ is independent of it.} With this, in the limit of enforcing the constraint strictly, the flow equation reads 
\begin{equation}
 \partial_t \Gamma_k \simeq \frac{1}{2} {\rm{STr}} \left[ \theta\left( 1 - \frac{\phi^2}{\phi_0^2}  \right) \frac{1}{\Gamma_k^{(2)} + R_k} \partial_t R_k \right] \, ,
\end{equation}
where we  have indicated that potentially boundary terms have to be treated carefully, due to the distributional character of the Heaviside distribution. In conclusion, an inequality constraint can  be implemented directly at the level of the Wetterich equation. The above form makes it clear that no polynomial truncation of the action in the field will be influenced by inequality constraints, with the exception of an expansion point directly on the boundary.

Let us finally discuss implications of this result on calculations using the background field method and the \FRG{}. In most calculations, a finite order expansion in the fluctuation field is performed. Such an expansion is insensitive to the Heaviside distribution in the flow, since its Taylor expansion around almost any configuration is trivial. Therefore, only calculations that include the full field dependence \cite{Knorr:2017mhu} are sensitive to inequality constraints.

We close with a word of caution by illustrating the case of an Ising model with the above constraint, in three dimensions. The unconstrained system features a single fixed-point  solution, namely the Wilson-Fisher fixed point, which is relevant for many condensed matter applications. Studying the differential equation for the scalar potential reveals that the condition of global existence of the solution singles out the Wilson-Fisher fixed point. Concretely, if $\phi_0$ in the above inequality constraint is chosen large enough, then the Wilson-Fisher fixed point is still the single solution to the fixed point equation. Conversely, if $\phi_0$ is smaller than the critical value, we expect infinitely many fixed point solutions to exist. This expectation springs from local existence theorems for differential equations. We conclude that an inequality constraint potentially changes the structure of the \RG{} flow dramatically.

\section{Summary and conclusions} \label{sec:conclusions}

In this work we discussed some details and problems in the construction of the metric configuration space for the gravitational path integral in both Euclidean and Lorentzian signature. Continuum approaches to quantum gravity based on this path integral typically make use of the background field formalism. The spacetime metric is parameterized in terms of a fluctuation field around a fixed, albeit arbitrary, background metric. The path integral is then written as a sum over all possible metric fluctuations. A key question is how to constrain the metric fluctuations such that the functional integral extends over spacetime configurations belonging to a subset of all possible metrics, \eg{}, metrics having a fixed signature and/or topology, without over-counting certain configurations or missing out on others.

A possible strategy to implement the constraint on the configuration space of the gravitational path integral consists of selecting suitable parameterizations for the metric fluctuations that naturally allow to reach only a subset of all possible metrics. Mathematically, this procedure provides a bijective map between the space of metrics of fixed signature and topology, and the space of real symmetric metric fluctuations, cf.~figure \ref{fig:illustration1}.
Concretely, we showed that for Euclidean spacetimes of fixed topology, there exist infinitely many bijective parameterizations, see, \eg{}, \eqref{eq:famparam}. Thus, restricting the signature of the metrics in a Euclidean path integral can be done within the standard background field method, simply by choosing an appropriate parameterization, as outlined in subsection \ref{sec:exEuclid}. Furthermore, in section \ref{sec:foliation} we found indications that the same parameterizations, together with the choice of a foliatable Euclidean background, is sufficient to implement a foliation structure for the full metric. Thus, restricting the path integral further to only include foliatable Euclidean manifolds does not pose additional constraints.

A more severe problem in the construction of the configuration space for quantum gravity arises when trying to restrict the set of all possible metrics to Lorentzian spacetimes only. Specifically, our discussion in section \ref{sec:Lorentzian} has highlighted the difficulty in combining the background field method with a path integral that includes Lorentzian metrics only. In particular, we have constructed an explicit example which shows that parameterizations that work for Euclidean spacetimes fail in the Lorentzian case. This could constitute a major challenge for the Asymptotic Safety program using \FRG{} techniques, which is currently explored almost exclusively in a Euclidean setting (see \cite{Manrique:2011jc, Fehre:2021eob, Banerjee:2022xvi, DAngelo:2022vsh} for exceptions).

To explore the Lorentzian setting, one might hope to obtain the results in Lorentzian signature by means of an analytic continuation of the Euclidean results, \ie{}, a Wick rotation. While in the case of quantum field theories on a flat background this procedure is expected to give the correct result, in the case of quantum field theories on a curved and fluctuating spacetime defining an analytical Wick rotation is an outstanding open problem~\cite{Baldazzi:2018mtl}. Moreover, an aspect of Wick rotation that is typically less explored is the question whether the two configuration spaces are in a one-to-one correspondence. Our results highlight that within the background field method this is not the case, since parameterizations of metric fluctuations that provide a bijective map between metric fluctuations $h_{\mu\nu}$ and the full metric in Euclidean spacetimes fail to do so in Lorentzian spacetimes. Therefore, an analytic continuation of the Euclidean results can -- at best -- give access to those Lorentzian metrics which are continuously connected to the flat Minkowski metric. 

Our results indicate that any path integral approach to quantum gravity, which is based on the metric, has to consider an explicit inequality constraint when exploring Lorentzian quantum gravity. In section \ref{sec:ineq-constraints} we discussed how to implement this inequality constraint within a simple scalar field theory and we showed how the \FRG{} equation is modified in the presence of a simple inequality constraint.

An interesting, alternative interpretation of our results is that quantum fluctuations of spacetime could also trigger signature changes. In this case, one should not limit the quantum-gravitational path integral to configurations having a specific signature. This is in line with the old idea \cite{Hartle:1983ai, Hawking:1983hj, Gibbons:1994cg} that the universe might have been a (static) Euclidean manifold, which then turned into the evolving Lorentzian universe which we observe today.
If instead one sticks on the idea that the spacetime signature should not fluctuate, an alternative setup to search for Asymptotic Safety in Lorentzian signature is worthwhile exploring. In \cite{Eichhorn:2017bwe, Eichhorn:2019xav}, it has been proposed that the causal set setup could be such a setting. Causal sets \cite{Bombelli:1987aa, Dowker:2013dog, Surya:2019ndm} can be viewed as discrete, causal networks of spacetime points. Both geometric as well as topological properties of a manifold can to a large extent be encoded in its causal order, see \cite{Sorkin:2018tbf} for recent examples.
The subset of causal sets known as sprinklings corresponds to discretizations of Lorentzian manifolds. If the discretization scale can be taken to zero at a higher order phase transition in the phase diagram of causal sets (spanned by the bare values of the couplings), then a universal continuum limit exists for the path integral over Lorentzian manifolds, such that Lorentzian Asymptotic Safety is realized. For studies of the phase diagram of causal sets in restricted configuration spaces, see \cite{Surya:2011du, Glaser:2017sbe, Glaser:2018jss, Cunningham:2019rob}. For related work in discrete networks for quantum gravity, see also \cite{Kelly:2018diy}. 

We close our discussion with a comment on degenerate metrics and spatial topology changes in view of our results, highlighting key differences between the Euclidean and Lorentzian cases. Keeping the signature of the full metric fixed implies that the metric cannot be degenerate. In fact, parameterizing the metric fluctuations by means of a map which satisfies Silverster's law of inertia, \eqref{eq:SilvLaw}, automatically excludes degenerate metrics, since the map $f$ has to be invertible.

In a Lorentzian universe, there are many examples of spacetimes which can undergo a variation (with time) of the topology of the spatial slices. An emblematic example is that of ``trousers spacetimes'', where the universe (the ``trunk'') branch off as a function of time and splits into two baby universes (the ``legs'' of the trousers). The metric of this spacetime is thus degenerate at the branch point. Using the background field method and a suitably implemented constraint on metric fluctuations, such a metric would be excluded by the ``sum over paths'' in the gravitational path integral.

Spacetimes exhibiting changes of topology can also be defined in the Euclidean. In this case, a topology change does not occur in time, but in space. A key example is that of Euclidean wormhole, where a ``spatial tube'' connects different (spatial) regions of the Euclidean manifold. At variance with the trousers spacetimes of the Lorentzian case, Euclidean wormholes do not feature degeneracies in the metric \cite{Giddings:1987cg, Hebecker:2018ofv}. Therefore, the corresponding metric would not be excluded by the (Euclidean) gravitational path integral, provided that the metric fluctuations are parameterized with a map satisfying \eqref{eq:SilvLaw}.

\section*{Acknowledgements}
The authors thank Astrid Eichhorn for fruitful collaboration in earlier stages, and for many discussions during various phases of the project. The authors also thank Bianca Dittrich for constructive comments on the manuscript. 
During parts of this work, A.~P.~was supported by the Alexander von Humboldt Foundation, and M.~S.~was supported by a scholarship of the German Academic Scholarship Foundation as well as by the DFG under grant no.~EI/1037-1. 
The authors acknowledge support by Perimeter Institute for Theoretical Physics. Research at Perimeter Institute is supported in part by the Government of Canada through the Department of Innovation, Science and Economic Development and by the Province of Ontario through the Ministry of Colleges and Universities. M.~S.~gratefully acknowledges extended hospitality at Syracuse University, and at CP3-Origins at the University of Southern Denmark during various stages of this work.

\appendix

\section{Proof that \texorpdfstring{$S^T \bar{g} S=\bar g S^2$}{STbargS=bargS2}}
\label{sec:Squared}

According to the Sylvester's law of inertia \cite{Sylvester:1852}, a symmetric square matrix $g_{\mu\nu}$ and another symmetric square matrix $\bar g_{\mu\nu}$ share the same number of positive, negative and zero eigenvalues if and only if they are related by the similarity transformation in \eqref{eq:SilvLaw}. In this Appendix we show that $S^T \bar{g} S=\bar g S^2$, provided that the matrix $S$ depends only on the combination $\hmat =\bar g^{-1} h$ and on the identity.

In order to prove that $S^T \bar{g} S=\bar g S^2$, it is sufficient to notice that if $S$ is a matrix that depends on $\hmat$ only, then it admits the representation \cite{Sylvester:1852}
	\begin{equation}
	S = S_0 \mathbbm 1 + S_1 \hmat + S_2 \hmat^2 + S_3 \hmat^3 \, ,
	\end{equation}
We can thus write
	\begin{equation}
	\begin{aligned}
		S^T \bar g S &= S^T (S_0 \bar g + S_1 h + S_2 h \bar g^{-1} h + S_3 h \bar g^{-1} h \bar g^{-1} h) \\
		&= S^T (S_0 \bar g + S_1 h + S_2 h \bar g^{-1} h + S_3 h \bar g^{-1} h \bar g^{-1} h)^T \\
		&= ((S_0 \bar g + S_1 h + S_2 h \bar g^{-1} h + S_3 h \bar g^{-1} h \bar g^{-1} h) S)^T \\
		&= (S_0 \bar g + S_1 h + S_2 h \bar g^{-1} h + S_3 h \bar g^{-1} h \bar g^{-1} h)S \\
		&= \bar g S^2 \, .
	\end{aligned}
	\end{equation}
In the first step we inserted $S$ and multiplied it by $\bar g$. In the second step, we used that the bracket is a symmetric matrix, so that we can take the transpose without changing anything. The third step combines the product of transposes to the transpose of the product in reverse order. We again realize that the bracket is a symmetric matrix, so we can remove the transpose. Finally, we notice that the bracket is nothing else than $\bar g S$, which completes the proof. This is non-trivial, and \emph{only} works because $S$ is a function of $\hmat$ (and the identity) only.

\bibliographystyle{JHEP}
\addcontentsline{toc}{section}{\refname}
\bibliography{references,manuals}

\end{document}